\documentclass{pasa}%

\usepackage{graphicx}

\title[The 1.4 GHz Cosmic Star Formation History at $z<1.3$]{The 1.4 GHz Cosmic Star Formation History at $z<1.3$}

\author[Upjohn et al.]{James E. Upjohn$^{1,2}$, Michael J. I. Brown$^{1,2}$, Andrew M. Hopkins$^3$ and Nicolas J. Bonne$^4$\thanks{michael.brown@monash.edu}
\affil{$^1$School of Physics and Astronomy, Monash University, Clayton, Victoria 3800, Australia}
\affil{$^2$Monash Centre for Astrophysics, Monash University, Clayton, Victoria, 3800, Australia}
\affil{$^3$Australian Astronomical Optics, Macquarie University, 105 Delhi Rd, North Ryde, NSW 2113, Australia}
\affil{$^4$Institute of Cosmology \& Gravitation, University of Portsmouth, Dennis Sciama Building, Portsmouth, PO1 3FX, UK}
}

\jid{PASA}
\doi{10.1017/pas.\the\year.xxx}
\jyear{\the\year}

\usepackage{aas_macros}
\usepackage{hyperref} 
\hypersetup{colorlinks,citecolor=blue,linkcolor=blue,urlcolor=blue}

\hypersetup{draft}

\begin{document}

\begin{frontmatter}
\maketitle

\begin{abstract}
We measure the cosmic star formation history out to $z=1.3$ using a sample of 918 radio-selected star forming galaxies within the 2 ${\rm deg^2}$ COSMOS field. To increase our sample size, we combine 1.4~GHz flux densities from the VLA-COSMOS catalogue with flux densities measured from the VLA-COSMOS radio continuum image at the positions of $I<26.5$ galaxies, enabling us to detect 1.4~GHz sources as faint as $40~{\rm \mu Jy}$. We find radio measurements of the cosmic star formation history are highly dependent on sample completeness and models used to extrapolate the faint end of the radio luminosity function. For our preferred model of the luminosity function, we find the star formation rate density increases from $0.019~M_\odot~{\rm yr}^{-1}~{\rm Mpc}^{-3}$ at $z \sim 0.225$ to $0.104~M_\odot~{\rm yr}^{-1}~{\rm Mpc}^{-3}$ at $z \sim 1.1$, which agrees to within $33\%$ of recent UV, IR and 3~GHz measurements of the cosmic star formation history. 
\end{abstract}

\end{frontmatter}

\section{INTRODUCTION }
\label{sec:introduction}

The star formation rate density (SFRD) quantifies the star formation rate (SFR) per unit volume at a given redshift, and has been measured at a variety of wavelengths (e.g. UV, IR, optical emission lines). Previous SFRD measurements at radio wavelengths are qualitatively similar to SFRDs measured with UV and IR data \citep[][and references therein]{madau2014}, although individual $z \sim 0$ radio luminosity functions can differ from each other by a factor of 3 at low radio powers \citep[e.g.][and references therein]{mauch2007}. All measurements of the cosmic star formation history (CSFH) at $z<3$ follow a broadly similar trend, with SFRD declining by a factor of roughly 20 since its peak at $z\sim 2.5$. For individual galaxies, SFRs determined with different indicators can often differ by factors of 2 or more, and inconsistencies as large as a factor of 10 are not uncommon \citep[e.g.][]{hopkins2003}. Consequently, achieving precise measurements of the CSFH with different SFR indicators requires large sample sizes to mitigate both cosmic variance and uncertainties on individual SFR measurements. At $z<1.5$ recent measurements of the CSFH agree with each other to within $\sim 0.2~{\rm dex}$ \citep[][and references therein]{madau2014}. 

By using radio data to measure the CSFH, it is possible to present a complementary view to the UV and IR surveys, which make up the bulk of high-$z$ SFRDs \citep{madau2014}. The advantage of radio observations is that they provide SFRs unbiased by dust but with the disadvantage of potentially including contamination from active galactic nuclei (AGNs). These AGNs have to be removed to measure a reliable radio SFRD, which can then be compared to SFRDs derived from other wavelengths and models.

The first attempts to calculate the CSFH from radio surveys used small samples of star forming (SF) galaxies, due to the relatively shallow data and/or small survey areas available at the time, and thus these studies had large uncertainties. One of the first measurements of the radio CSFH, by \citet{haarsma2000}, found that the SFRD has declined by an order of magnitude since $z\sim 1.6$, but their best sample contained just 37 radio sources. Over the subsequent decade the radio sample sizes increased, with \citet{seymour2008} measuring the CSFH with 269 $z<3$ star forming galaxies and \citet{smolcic2009} using 340 $z<1.3$ star forming galaxies in the COSMOS field. The most recent study of the radio CSFH, by \citet{novak2017}, uses 5806 $z<5.7$ star forming galaxies with 3~GHz flux densities from the VLA-COSMOS 3~GHz Large Project. Despite the large expansion in the sample sizes over the past two decades, the radio luminosity function and CSFH can still be adequately modelled with pure luminosity evolution \citep{novak2017}, as previously found by \citet{hopkins2004}.

Previous measurements of the radio CSFH have largely used blind radio survey source catalogues. Such catalogues require conservative signal-to-noise (S/N) thresholds to avoid being swamped by large numbers of spurious sources. For example, searching images containing a total of 250 million pixels with a S/N threshold of 4 could produce as many as $\sim 8000$ false detections. As the number of spurious sources scales with the number of pixels, lower S/N thresholds can be used if one limits detections and measurements to positions of optically identified galaxies. This approach may be particularly useful for studies exploiting the new generation of deep radio continuum surveys in combination with optical galaxy surveys \citep[e.g.,][]{norris2011,dacunha2017}, enabling significant expansion of sample sizes. In this letter we utilise this approach to detect sources down to a signal-to-noise of 3 and to measure the CSFH.

We present the $z<1.3$ CSFH measured by combining the COSMOS $I<26.5$ object catalogue with the $1.4~{\rm GHz}$ VLA-COSMOS survey. We exploit a combination of catalogued radio sources and new flux density measurements at the positions of optical galaxies to reliably measure $1.4~{\rm GHz}$ flux densities as faint as $S_{1.4}=40~{\rm \mu Jy}$. We use AB magnitudes throughout, the \citet{salpeter1955} initial mass function (IMF), and adopt a flat nine-year WMAP cosmology with $H_0=69.32~{\rm km~s~Mpc^{-1}}$, $\Omega_M=0.29$ and $\Omega_\Lambda=0.71$ \citep{bennett2013}. We assume the radio synchrotron spectrum is approximated by $S_\nu \propto \nu^{-0.7}$.

\section{Sample}
\label{sec:sample}

We selected our sample using a combination of COSMOS optical and radio catalogues. The main COSMOS optical catalogue is from \citet{ilbert2015}, and it contains over 2 million sources with $I<26.5$. This was cross-matched to both the \citet{lilly2007} spectroscopic redshift catalogue and the COSMOS2015 photometric redshift catalogue \citep{laigle2016}. When available, we use spectroscopic redshifts in preference to photometric redshifts. 

Our principal radio catalogue is from the 1.4 GHz VLA-COSMOS survey \citep{schinnerer2010}, which imaged $2~{\rm deg}^2$ centred on 10:00:28.60~+02:12:21.00 with an angular resolution of $2.5^{\prime\prime}$. In the central $50^\prime\times50^\prime$ the depth is $S_{1.4}={\rm 12~\mu Jy}$ RMS, increasing to an RMS of $S_{1.4}={\rm 34~\mu Jy}$ at the edges. The accompanying source catalogue includes sources with ${\rm S/N} >4\sigma$. The VLA-COSMOS deep radio catalogue of \citet{schinnerer2010} does not include the suite of COSMOS multiwavelength data, so magnitudes and redshifts were obtained by finding the nearest cross-matches in the merged optical, spectroscopic redshift and photometric redshift catalogue. 

If we exclude multicomponent sources from the \citet{schinnerer2010} radio catalogue, as they are likely to be AGNs, then we have 2732 radio sources. When we match these 2732 sources to the optical catalogue, we find that just 233 (9\%) of the radio catalogue sources lack optical counterparts. For radio sources fainter than $100~{\rm \mu Jy}$, we find the percentage without optical counterparts rises to just 11\%. Using the \citet{brown2017} nearby galaxy sample, we find star forming galaxies have $i$-band to $1.4~{\rm GHz}$ flux density ratios that range from 0.1 to 30, so we expect low redshift $S_{1.4} = 40~{\rm \mu Jy}$ star forming galaxies to be brighter than $i=22.4$. At $z=1.15$ the $i$-band and $1.4~{\rm GHz}$ correspond to $u$-band and $3.0~{\rm GHz}$ respectively, and we find the \citet{brown2017} nearby galaxy sample $u$-band to $3.0~{\rm GHz}$ flux density ratios range from 0.05 to 20, so we expect $z\sim 1.15$ $S_{1.4} = 40~{\rm \mu Jy}$ star forming galaxies to be brighter than $i=23.2$. We thus conclude that the bulk of the radio sources without $i<26.5$ optical counterparts are AGNs or $z>1.3$ galaxies.

Our sample uses a combination of the \citet{ilbert2015} optical catalogue and the VLA-COSMOS deep catalogue of \citet{schinnerer2010}, supplemented by $1.4~{\rm GHz}$ flux densities measured at the positions of the optical galaxies. For every optical galaxy we measured a $1.4~{\rm GHz}$ flux density per beam using the pixel in the VLA-COSMOS images that corresponds to the optical position, although we only use this flux density if a match to the \citet{schinnerer2010} catalogue is unavailable. (This assumes we are mostly observing radio point sources, which is a reasonable approximation for star forming galaxies across the bulk of our redshift range, as $2.5^{\prime\prime}$ corresponds to 8.3~kpc at $z=0.2$.) For the subsequent analyses we retain galaxies with $1.4~{\rm GHz}$ ${\rm S/N} >3\sigma$ and $S_{1.4} \geq 40~{\rm \mu Jy}$ measured in the \citet{schinnerer2010} catalogue and/or measured directly from the VLA-COSMOS images.
 
\begin{figure}[tbp]
\begin{center}
\resizebox{3.25in}{!}{\includegraphics{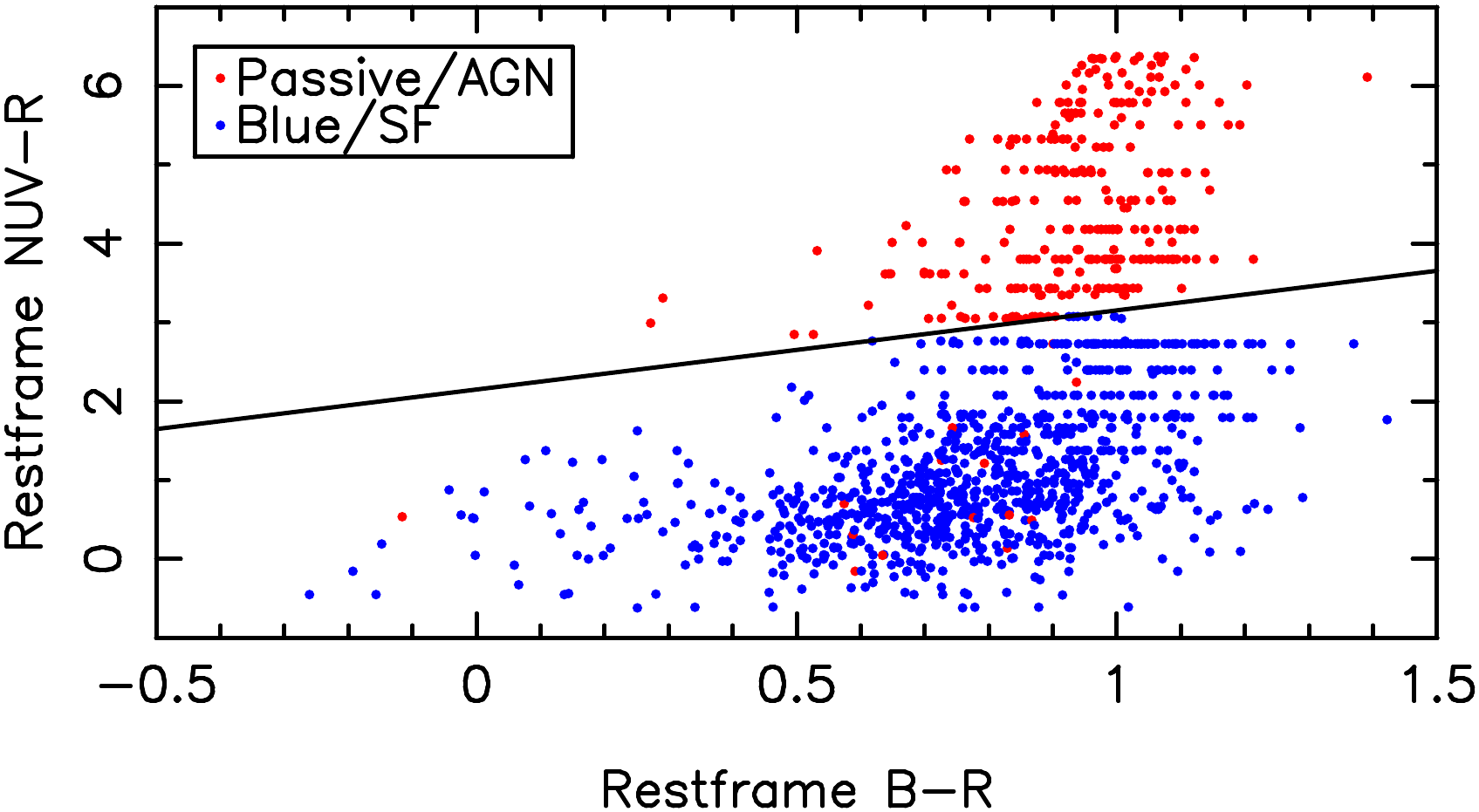}}
\caption{The restframe colours of our final sample of 1218 radio selected galaxies, illustrating the loci of SF and passive galaxies. We use the empirical criterion shown above in combination with a modified \citet{stern2005} mid-infrared wedge to split the sample into 918 star forming galaxies and 300 AGNs. (Some quantization and aliasing of the absolute magnitudes is an artifact from the COSMOS2015 photometric redshift catalogue.)} 
\label{fig:colour}
\end{center}
\end{figure}

To measure the radio CSFH we selected a sample of $z<1.3$ star forming galaxies with radio flux densities above $3\sigma$, using regions that have two or more VLA pointings in the VLA-COSMOS survey \citep[Fig. 2 from][]{schinnerer2010}. Radio sources with $S_{1.4}>50~{\rm mJy}$ that cause artifacts in the VLA-COSMOS deep image were identified and all the galaxies in these regions were removed. This reduces our central area, where the depth is $S_{1.4}=40~{\rm \mu Jy}$, to $0.72~{\rm deg}^2$, while our (shallower) total area is $0.96~{\rm deg}^2$. We also visually cross checked close pairs of radio sources and removed instances of single sources being spuriously identified as multiple objects. We then split our sample into star forming galaxies and AGNs using the passive galaxy colour criterion illustrated in Figure~\ref{fig:colour} and a modified \citet{stern2005} infrared AGN selection criterion (that does not cut into the galaxy locus). This leaves us with a sample containing 918 $z<1.3$ star forming galaxies, of which 374 were not contained within the \citet{schinnerer2010} VLA-COSMOS deep catalogue. Of the 918 $z<1.3$ star forming galaxies with $S_{1.4}>40~{\rm \mu Jy}$, 883 (96\%) are brighter than $I=25$, consistent with incompleteness due to our $I=26.5$ magnitude limit being negligible. 

Photometric redshift uncertainties for $23<I<24$ $z\sim 1$ star forming galaxies are $4.4\%$ \citep{laigle2016}, which propagate through to individual radio powers having uncertainties of 9\% (while brighter star forming galaxies have smaller uncertainties). We use spectroscopic redshifts in preference to photometric redshifts when possible, and they are available for 37\% of the $I<22$ star forming galaxies. 

\section{The 1.4 GHz Radio Luminosity Function}

We have measured the radio luminosity function for four different redshift bins using the standard $1/V_{max}$ method \citep{schmidt1968}. To achieve this, our sample was split into four redshift ranges ($0.1\leq\rm{z}<0.35,\ 0.35\leq\rm{z}<0.6,\ 0.6\leq\rm{z}<0.9\ and\ 0.9\leq\rm{z}<1.3$) and then into bins of $1/3~{\rm dex}$ in luminosity. Luminosity limits and survey volumes are constrained by the flux density limits of $S_{1.4}=40~{\rm \mu Jy}$ for the central region and $S_{1.4}=60~{\rm \mu Jy}$ for the outer $350^{\prime\prime}$. To keep the sample complete these flux limits were taken into account when calculating the redshift range, effective area and resulting $V_{max}$ for each bin. As a result, of the 918 star forming galaxies in our sample, 525 meet our area and redshift dependent luminosity criteria. 


\begin{table*} 
\caption{1.4 GHz radio luminosity functions for VLA-COSMOS star forming galaxies.\label{tab: sf luminosity}}
\begin{center}
\begin{tabular*}{\textwidth}{ccccc}
\hline \hline
Redshift &  Luminosity range                             & Number of & Volume                   & LF                                         \\
range     &  $\rm{log}_{10} (\rm{W~Hz^{-1}})$  & galaxies     &  (${\rm Mpc^{-3}}$) & (${\rm Mpc^{-3}~dex^{-1}}$) \\
\hline 
0.10$\leq$$z$$<$0.14 & 21.33-21.67 & 3 & $7.42 \times 10^{3}$ & 1.21 $\pm^{1.36}_{0.57}\times 10^{-3}$   \\ 
0.10$\leq$$z$$<$0.20 & 21.67-22.00 & 8 & $3.38 \times 10^{4}$ & 7.11 $\pm^{3.86}_{2.23}\times 10^{-4}$   \\ 
0.10$\leq$$z$$<$0.28 & 22.00-22.33 & 38 & $9.89 \times 10^{4}$ & 1.15 $\pm^{0.23}_{0.18}\times 10^{-3}$   \\ 
0.10$\leq$$z$$<$0.35 & 22.33-22.67 & 43 & $2.05 \times 10^{5}$ & 6.29 $\pm^{1.17}_{0.91}\times 10^{-4}$   \\ 
0.10$\leq$$z$$<$0.35 & 22.67-23.00 & 17 & $2.05 \times 10^{5}$ & 2.49 $\pm^{0.82}_{0.56}\times 10^{-4}$   \\ 
0.10$\leq$$z$$<$0.35 & 23.00-23.33 & 2 & $2.05 \times 10^{5}$ & 2.92 $\pm^{4.51}_{1.62}\times 10^{-5}$   \\ 
\hline \\ 
0.35$\leq$$z$$<$0.40 & 22.33-22.67 & 23 & $7.02 \times 10^{4}$ & 9.84 $\pm^{2.67}_{1.91}\times 10^{-4}$   \\ 
0.35$\leq$$z$$<$0.56 & 22.67-23.00 & 58 & $4.72 \times 10^{5}$ & 3.69 $\pm^{0.57}_{0.46}\times 10^{-4}$   \\ 
0.35$\leq$$z$$<$0.60 & 23.00-23.33 & 27 & $6.38 \times 10^{5}$ & 1.27 $\pm^{0.31}_{0.23}\times 10^{-4}$   \\ 
0.35$\leq$$z$$<$0.60 & 23.33-23.67 & 4 & $6.38 \times 10^{5}$ & 1.88 $\pm^{1.69}_{0.79}\times 10^{-5}$   \\ 
0.35$\leq$$z$$<$0.60 & 23.67-24.00 & 1 & $6.38 \times 10^{5}$ & 4.70 $\pm^{13.05}_{3.31}\times 10^{-6}$   \\ 
\hline \\ 
0.60$\leq$$z$$<$0.78 & 23.00-23.33 & 86 & $7.03 \times 10^{5}$ & 3.67 $\pm^{0.45}_{0.38}\times 10^{-4}$   \\ 
0.60$\leq$$z$$<$0.90 & 23.33-23.67 & 39 & $1.38 \times 10^{6}$ & 8.50 $\pm^{1.67}_{1.29}\times 10^{-5}$   \\ 
0.60$\leq$$z$$<$0.90 & 23.67-24.00 & 15 & $1.38 \times 10^{6}$ & 3.27 $\pm^{1.16}_{0.77}\times 10^{-5}$   \\ 
0.60$\leq$$z$$<$0.90 & 24.00-24.33 & 4 & $1.38 \times 10^{6}$ & 8.72 $\pm^{7.82}_{3.67}\times 10^{-6}$   \\ 
\hline \\ 
0.90$\leq$$z$$<$1.09 & 23.33-23.67 & 72 & $1.10 \times 10^{6}$ & 1.96 $\pm^{0.27}_{0.22}\times 10^{-4}$   \\ 
0.90$\leq$$z$$<$1.30 & 23.67-24.00 & 62 & $2.66 \times 10^{6}$ & 7.00 $\pm^{1.05}_{0.85}\times 10^{-5}$   \\ 
0.90$\leq$$z$$<$1.30 & 24.00-24.33 & 18 & $2.66 \times 10^{6}$ & 2.03 $\pm^{0.64}_{0.44}\times 10^{-5}$   \\ 
0.90$\leq$$z$$<$1.30 & 24.33-24.67 & 5 & $2.66 \times 10^{6}$ & 5.64 $\pm^{4.29}_{2.17}\times 10^{-6}$   \\ 
\hline
\end{tabular*}
\end{center}
\tabnote{$^a$This volume includes the outer region with the shallower flux density limit.}
\end{table*}

\begin{figure*}[tbp]
\begin{center}
\resizebox{3.25in}{!}{\includegraphics{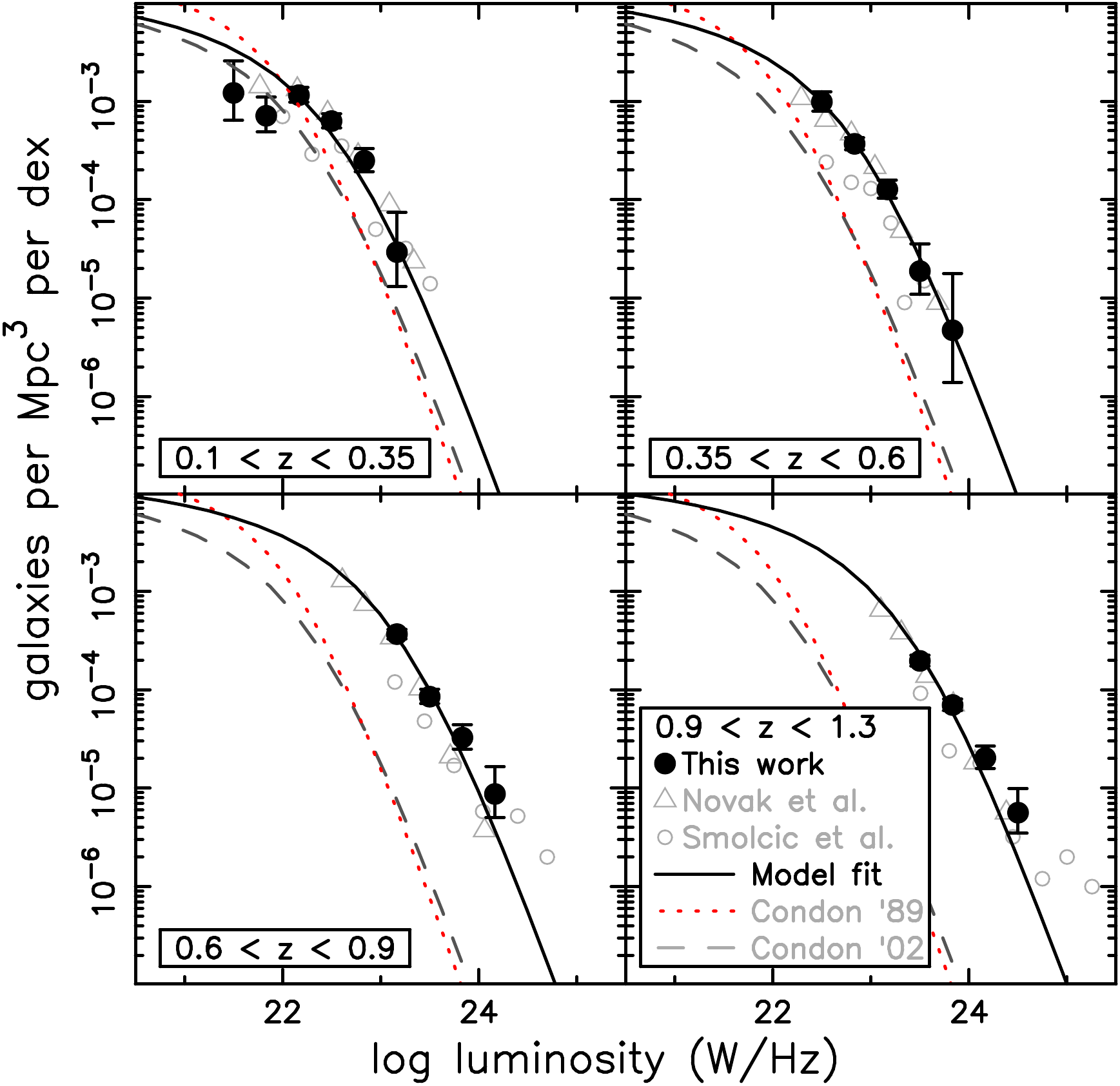}} 
\caption{1.4 GHz luminosity functions of star forming galaxies. The dashed line is the \citet{condon2002} local luminosity function, while the solid line is this function fitted to our data under the assumption of pure luminosity evolution. The red dotted line is the deprecated \citet{condon1989} local luminosity function, which overestimates the number of galaxies at low radio luminosities. We measure a higher space density of low luminosity galaxies than \citet{smolcic2009}, but are in agreement with the deeper 3 GHz measurements of \citet{novak2017}.}
\label{fig:lumfunc}
\end{center}
\end{figure*}

In Figure~\ref{fig:lumfunc} and Table~\ref{tab: sf luminosity} we present the 1.4 GHz luminosity functions for our four redshift bins, including uncertainties determined using Poisson statistics \citep[e.g.,][]{gehrels1986}. However, we caution that for the inner region of the sample the $0.10<z<0.14$ and $0.10<z<0.20$ redshift slices have cosmic variance uncertainties of $74\%$ and $43\%$ respectively \citep[determined with the method of ][]{driver2010}. For comparison in Figure~\ref{fig:lumfunc} we also show the VLA-COSMOS 1.4 and 3.0~GHz luminosity functions previously measured by \citet{smolcic2009} and \citet{novak2017}, adopting a spectral index of -0.7 for frequency conversions. Compared to the \citet{smolcic2009} $1.4~{\rm GHz}$ luminosity functions, we find higher space densities at all but the very highest radio luminosities. This is expected, as 41\% of our star forming galaxies were not included in the VLA-COSMOS radio source catalogue of \citet{schinnerer2010}. We find better agreement with the recent luminosity functions of \citet{novak2017}, who also measure higher space densities than \citet{smolcic2009}, using a sample selected from 3~GHz imaging of the COSMOS field (with an RMS of $S_{3.0}=2.3~{\rm \mu Jy}$ per beam).

To model the luminosity function we use the models of \citet{condon1989} and \citet{condon2002} with pure luminosity evolution added. Our preferred model is \citet{condon2002}, which is based on the luminosity function determined with 1966  $\rm{m_p<14.5}$ galaxies from the Uppsala General Catalogue \citep{nilson1973} with counterparts in the NRAO VLA Sky Survey \citep{condon1998}. However, we also make use of the \citet{condon1989} model as it has been used by other studies during the past decade, including \citet{smolcic2009}, and it illustrates the impact of model assumptions on our results and those from the literature. The dashed lines in Figure~\ref{fig:lumfunc} illustrate the $z=0$ \citet{condon2002} luminosity function while the solid lines show fits of the \citet{condon2002} model to our data under the assumption of pure luminosity evolution. When we approximate the observed luminosity evolution with $L(z)\propto (1+z)^Q$, we find $Q=3.10\pm 0.06$. 

\section{The Cosmic Star Formation History}

\begin{figure*}[tbp]
\begin{center}
\resizebox{3.25in}{!}{\includegraphics{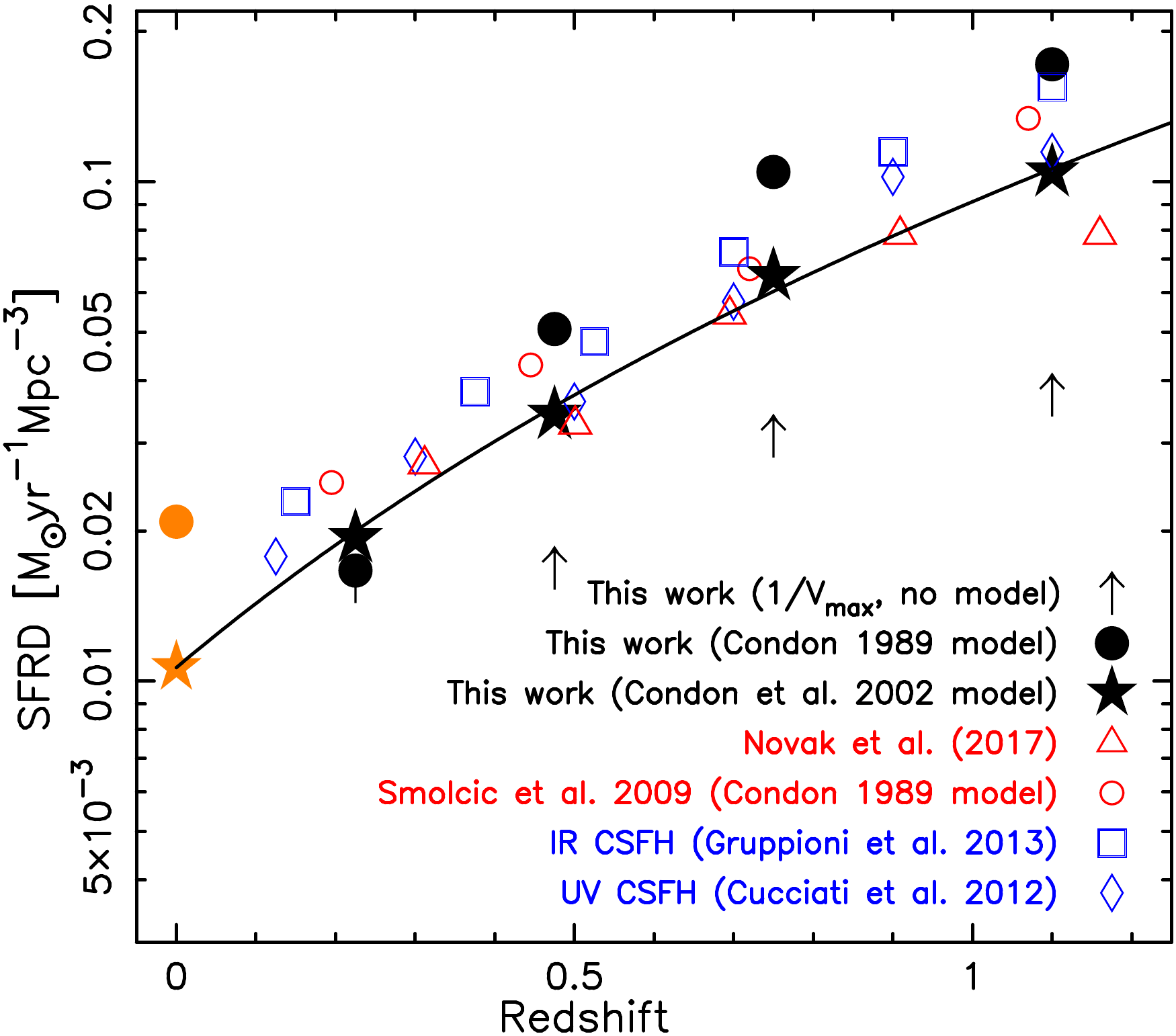}} 
\caption{Our measurements of the CSFH, along with UV, IR, 3 GHz and 1.4 GHz measurements from the recent literature \citep{smolcic2009,cucciati2012,gruppioni2013,novak2017}. For clarity we do not show random uncertainties as they are comparable to or smaller than the symbols. Our SFRD measurements include lower limits determined with the $1/V_{max}$ method, estimates determined with fits of the deprecated \citet{condon1989} model and estimates determined with fits of the preferred \citet{condon2002} model. Measurements of the $z=0$ SFRD, determined with the radio luminosity functions of \citet{condon1989} and \citet{condon2002}, are shown with the orange symbols. Our best measurement of CSFH is in broad agreement with recent UV, IR and 3 GHz results, and can be extrapolated to agree with the $z=0$ radio luminosity function of \citet{condon2002}. Pure luminosity evolution, shown with the black line, approximates our data well and indicates the SFRD has declined by a factor of ten since $z=1.1$.}
\label{fig:csfr}
\end{center}
\end{figure*}

To measure the SFRD and CSFH, we first determined each galaxy's SFR using the 1.4 GHz radio luminosity and the calibration of \citet{bell2003}:
\begin{equation}
\psi[M_{\odot}yr^{-1}]=\left\{\begin{array}{cc}  
5.52\times10^{-22}L_{1.4{\rm GHz}}, & L_{1.4{\rm GHz}}>L_c\\
\frac{5.52\times10^{-22}}{0.1+0.9(L_{1.4{\rm GHz}}/L_c)^{0.3}}L_{1.4{\rm GHz}}, & L_{1.4{\rm GHz}}\leq L_c
\end{array}\right.
\label{eq:sfr}
\end{equation}
where $L_c=6.4\times10^{21}~{\rm W~Hz^{-1}}$. This SFR calibration has been used by both \citet{smolcic2009} and \citet{novak2017}, and is comparable to recent calibrations by \citet{boselli2015} and \citet{brown2017} for $<L^*$ galaxies. We then used the $1/V_{max}$ method \citep{schmidt1968}, with the SFR  divided by the volume in which the galaxy could be detected, to find the SFRD contribution for each galaxy. The sum of each detected galaxy's contribution to the SFRD gives a lower limit to the SFRD, as it does not include star forming galaxies fainter than our radio flux limits. These lower limits for the CSFH are shown with arrows in Figure~\ref{fig:csfr}.


To obtain an SFRD that includes galaxies fainter that our flux density limits, we have modelled our data with the \citet{condon1989} and \citet{condon2002} radio luminosity function models with the addition of pure luminosity evolution where $L(z) \propto(1+z)^Q$. We choose to not use the newer radio luminosity function model of \citet{mauch2007} as their sample is incomplete at low luminosities due to the surface brightness limits of the 2MASS survey. While the \citet{condon1989} model has been used by some studies, we find it is a relatively poor fit to our data and it overestimates the $z\sim 0$ SFRD relative to recent radio luminosity functions \citep{condon2002,mauch2007,mao2012}. Our preferred model is thus \citet{condon2002}, which approximates our data well, as illustrated by Figure~\ref{fig:lumfunc}. 


\begin{table*} 
\caption{The radio SFRDs for the COSMOS field.\label{tab:csfr}}
\small
\begin{center}
\begin{tabular*}{\textwidth}{ccccc}
\hline \hline
Redshift & \multicolumn{2}{c}{This work - SFRD}                                                     & \multicolumn{2}{c}{\citet{smolcic2009} - SFRD}                                      \\
Range   &  \citet{condon2002} model              &  \citet{condon1989} model             & \citet{sadler2002} model               & \citet{condon1989} model             \\
              &  (${\rm M_\odot yr^{-1}Mpc^{-3}}$) & (${\rm M_\odot yr^{-1}Mpc^{-3}}$) & (${\rm M_\odot yr^{-1}Mpc^{-3}}$) & (${\rm M_\odot yr^{-1}Mpc^{-3}}$) \\
\hline
 $ 0.10 \leq z < 0.35 $ &  $ 0.019 \pm^{0.002}_{0.002} $ & $ 0.017 \pm^{0.002}_{0.002}$ & $ 0.023\pm^{0.002}_{0.002} $ & $ 0.025\pm^{0.001}_{0.001} $ \\
 $ 0.35 \leq z < 0.60 $ &  $ 0.034 \pm^{0.003}_{0.003} $ & $ 0.051 \pm^{0.003}_{0.003}$ & $ 0.032\pm^{0.003}_{0.002} $ & $ 0.043\pm^{0.003}_{0.003} $ \\
 $ 0.60 \leq z < 0.90 $ &  $ 0.065 \pm^{0.004}_{0.004} $ & $ 0.105 \pm^{0.005}_{0.005}$ & $ 0.048\pm^{0.003}_{0.004} $ & $ 0.067\pm^{0.003}_{0.003} $ \\
 $ 0.90 \leq z < 1.30 $ &  $ 0.104 \pm^{0.005}_{0.006} $ & $ 0.172 \pm^{0.007}_{0.008}$ & $ 0.088\pm^{0.005}_{0.005} $ & $ 0.134\pm^{0.010}_{0.009} $ \\

\hline
\end{tabular*}
\end{center}
\tabnote{$^a$This volume includes the outer region with the shallower flux density limit.}
\end{table*}

We determined SFRDs by integrating $\int\psi(L)\Phi_{B}(L)dL$, where $\psi(L)$ is the star formation rate from Equation \ref{eq:sfr} and $\Phi_B(L)$ is the radio luminosity function model. The resulting SFRDs and (for comparison) those from \citet{smolcic2009} are provided in Table~\ref{tab:csfr} and Figure~\ref{fig:csfr}. As Figure~\ref{fig:csfr} illustrates, our SFRD measurements depend on the luminosity function model, with the impact increasing with increasing redshift. At $0.10 \leq z < 0.35$ our SFRD measurements have little dependence on the luminosity function model, but at $0.90 \leq z < 1.30$ the \citet{condon1989} model provides an estimate of the SFRD that is 65\% higher than the estimate provided using the \citet{condon2002} model. 

As expected, we measure a systematically higher CSFH than \citet{smolcic2009} when the same luminosity function model is used. However, it should be noted that \citet{smolcic2009} does agree well with the literature when the \citet{condon1989} model of the luminosity function is used, as this model's overestimate of the number of faint sources compensates for sample incompleteness. Our SFRD measurements, using the preferred \citet{condon2002} model, agree well with recent UV, IR and 3~GHz studies \citep{cucciati2012,gruppioni2013,novak2017}, with the possible exception of $0.90 \leq z < 1.30$, where we measured a higher SFRD than \citet{novak2017}. Using the preferred \citet{condon2002} model. We find the SFRD can be approximated by $0.0109\times (1+z)^{3.10\pm 0.06} {\rm ~M_\odot~yr^{-1}~Mpc^{-3}}$, which implies SFRD has declined by a factor of ten from $0.90 \leq z < 1.30$ to $z\sim 0$. 

\section{Discussion}


Given both we and \citet{smolcic2009} measure the $1.4~{\rm GHz}$ luminosity function and CSFH using the COSMOS field, it is useful to explore why our results differ significantly from those of \citet{smolcic2009}. First, there are some differences in the radio catalogues used, with \citet{smolcic2009} using the \citet{schinnerer2007} catalogue whereas we are using the deeper \citet{schinnerer2010} catalogue. The most significant differences, however, arise from our addition of new flux densities measured at the positions of optical galaxies allowing the use of a lower S/N threshold. \citet{smolcic2009} imposed ${\rm S/N}>5\sigma$ and $S_{1.4} \geq 52.5~{\rm \mu Jy}$, whereas we imposed ${\rm S/N}>3\sigma$ and $S_{1.4} \geq 40.0~{\rm \mu Jy}$. Our sample contains 308 star forming galaxies with $40.0~{\rm \mu Jy} \leq S_{1.4} < 52.5~{\rm \mu Jy}$.

Given the large differences in sample size, are the radio flux densities measured at optical galaxy positions valid? We find flux densities (per beam) measured at optical galaxy positions are typically $\sim 90\%$ of the catalogued flux densities (per beam) for sources with $S_{1.4} \lesssim 100 ~{\rm \mu Jy}$. Visual inspection of the $1.4~{\rm GHz}$ VLA-COSMOS image shows that uncatalogued radio sources with ${\rm S/N}>3$ appear to be real radio sources rather than artifacts. This includes a limited number of $S_{1.4} \sim 100~{\rm \mu Jy}$ sources, which are significantly brighter than the notional limits of the VLA-COSMOS catalogs \citep{schinnerer2007,schinnerer2010}, which suggests there are (rare) failures of source detection algorithms on relatively bright sources (whose flux densities can otherwise be accurately measured). 

A final cross check of the validity of our approach is by comparison to previous literature. Our results agree well with recent measurements of the CSFH using UV, IR and $3~{\rm GHz}$ data \citep[e.g.,][]{cucciati2012,gruppioni2013,novak2017}. We find the evolution of the SFRD at $z<1.3$, determined by fitting our SFRD measurements and \citet{condon2002}, is approximated by $0.0109\times (1+z)^{3.10\pm 0.06} {\rm ~M_\odot~yr^{-1}~Mpc^{-3}}$. For comparison, \citet{novak2017} find radio luminosity function can be approximated by pure luminosity evolution where $L(z) \propto (1+z)^{(3.16\pm 0.2 - (0.32 \pm 0.07)z)}$, which is similar to our results at low redshift but does predict less evolution at $z>1$. Other measurements of the CSFH at $z<1$ are broadly similar \citep[e.g.,][]{hopkins2004}. Given the results of our cross checks and consistency with the literature, we conclude our approach to measuring radio luminosity function is valid. 

\section{Conclusions}

We have measured the CSFH at $z<1.3$ of 918 star forming galaxies in the COSMOS field with $1.4~{\rm GHz}$ radio continuum flux densities of $S_{1.4}>40~{\rm \mu Jy}$. We find that just 9\% of the radio sources from the VLA-COSMOS Large Project radio catalogue do not have optical counterparts down to a magnitude limit of $I=26.5$. We have also increased the size and depth of our sample by using radio flux densities measured at the positions of $I<26.5$ galaxies selected from the COSMOS optical catalogue. 


We find that our radio measurements of the CSFH, and those from the prior literature, have increasing dependence on the LF model with increasing redshift. For example, for our high redshift bin the deprecated \citet{condon1989} model produces SFRD measurements that are 65\% higher than those produced by the \citet{condon2002} model. We find that pure luminosity evolution of the \citet{condon2002} model approximates our data well, and the resulting SFRDs are in broad agreement with recent UV, IR and 3~GHz measurements. Using our preferred model and 525 $z<1.3$ star forming galaxies, we find that the SFRD is approximated by $0.0109\times (1+z)^{3.10\pm 0.06} {\rm ~M_\odot~yr^{-1}~Mpc^{-3}}$ and that it has declined by a factor of ten since $z=1.1$.

\section{Acknowledgements}

We are grateful to Olivier Ilbert and Peter Capak for  providing preliminary optical galaxy and photometric redshift catalogues during the development of this paper. MJIB acknowledges financial support from The Australian Research Council (FT100100280), the Monash Research Accelerator Program (MRA), the Monash Outside Studies Programme (OSP) and the University of Cambridge. Part of this work was undertaken while MJIB was on OSP (sabbatical) leave at the University of Cambridge, Swinburne University and the University of Melbourne. The National Radio Astronomy Observatory is a facility of the National Science Foundation operated under cooperative agreement by Associated Universities, Inc.

\bibliographystyle{pasa-mnras}
\bibliography{upjohn_radio}

\end{document}